\documentclass[useAMS,usenatbib,referee]{mn2e}

\def\hone{{\sc Hi}}
\def\msunpyr{\rm M_{\sun}\, yr^{-1}}
\def\psqcm{\rm cm^{-2}}
\def\ie{{i.e.}}
\def\eg{{e.g.}}

\newlength{\numberwidth}
\newlength{\dotwidth}
\title[Foreground Hydrogen over cooling flows]{Uniformity of
foreground Galactic neutral Hydrogen over cooling flow clusters}
\author[D. G. Barnes and P. E. J. Nulsen]{D. G. Barnes$^{1}$\thanks{E-mail:
dbarnes@isis.ph.unimelb.edu.au} and P. E. J. Nulsen$^{2}$\\
$^{1}$School of Physics, University of Melbourne, Vic 3010, Australia\\
$^{2}$Engineering Physics, University of Wollongong, NSW 2522,
Australia}
\begin{document}

\date{2002 July 9}

\pagerange{\pageref{firstpage}--\pageref{lastpage}} \pubyear{2002}

\maketitle

\label{firstpage}

\begin{abstract}
  Radio maps at 21 cm of foreground neutral hydrogen over three
  cooling flow clusters of galaxies show that the foreground gas is
  uniform on scales from $\sim 1$ -- 10 arcmin.  Five sigma
  limits on fluctuations in the foreground column density
  for Abell 3581, Sersic 159-03 and Abell 2597 are
  $6.2\times10^{19}$, $7.8\times10^{19}$ and $4.3\times10^{19} \,
  \psqcm$, or $14$, $43$ and $17$ percent of the mean foreground column
  density in the region of the system, respectively.  Fluctuations in
  the column density of neutral gas in the Galaxy are unlikely to
  account for any excesses of photoelectric absorption in these or
  other cooling flow clusters.  Fluctuations in the foreground neutral
  gas on arcminute scales are also unlikely to be the cause of excess
  EUV and soft X-ray emission from clusters.
\end{abstract}


\section{Introduction}

Data from {\it Chandra\/} and {\it XMM-Newton\/} have caused radical
revisions to the X-ray view of cooling flows in clusters of galaxies
\citep{f94}.  Where data from X-ray missions such as {\it ROSAT} and
{\it ASCA} were interpreted as showing that substantially more than
$100 \, \msunpyr$ of cooled gas is being deposited in the
central regions of many clusters (\citealt*{wjf}; \citealt{pfe, afj}),
{\it Chandra\/} and {\it XMM-Newton\/} data are inconsistent with such
high cooling rates \citep[e.g.][]{dnm}, and show little spectroscopic
evidence for gas cooling below temperatures of about 1 keV
\citep[e.g.][]{ppk,tkp}.

The cooling time of the gas at the centre of a cooling flow cluster is
typically only a few times $10^8\rm\, yr$, so that substantial heat
input is required to prevent large amounts of gas from cooling below 1
keV.  The rapidly growing list of cavities found in the X-ray emitting
gas associated with radio lobes emanating from the nucleus of the
central galaxy \citep[e.g.][]{mwn,fse,bsm} suggests strongly that
radio outbursts from the active nucleus of the central galaxy provide
the required heating \citep{csf,ndm}.

Many issues remain to be settled in the emerging picture of cluster
cooling flows \citep[e.g.][]{fmn}.  In particular, the presence of
cool gas in a wide variety of forms \citep[e.g.][]{cae,dmv,e01,ewj} and star
formation (e.g. \citealt*{jfn}; \citealt{mo92,cae,mkt}) in many of
these systems indicates that some of the X-ray emitting intracluster
gas does cool to low temperatures.  This is supported by the detection
of some X-ray emitting gas below 1 keV in at least one system
\citep{kft}.  However, it is yet to be determined how the cooled gas
arises.  It might be deposited in brief episodes every few times
$10^8\rm\, yr$, as the central gas radiates the last of its thermal
energy, cools to low temperatures and triggers a nuclear outburst that
chokes off further cooling.  Alternatively, it may be deposited as a
continuous thermally unstable trickle \citep[e.g.][]{ocd}.  It has
also been proposed that at least some of the cool gas did not
originate as intracluster gas, but was brought to the cluster centre
with infalling galaxies \citep{m97}.  As well, the state of the bulk
of the cool gas at the centres of cooling flows is disputed
(\citealt*{ffj}; \citealt{vd95}).

In this paper we discuss maps of Galactic foreground \hone\ emission
over cooling flow clusters.  This work was originally prompted by
measurements of large excess photoelectric absorption in cooling flow
clusters \citep{wfj,af97,a00}, implying very large reservoirs of
cooled gas within them.  The aim was to determine if arcminute scale
fluctuations in foreground \hone\ column density could account for the
excess absorption (\citealt*{djf} proposed that variations in
foreground dust column density could have this effect).  However, {\it
  Chandra} and {\it XMM-Newton} observations now suggest that the
excess photoelectric absorption was largely an artifact of models in
which large quantities of gas were assumed to cool from the ambient
temperature of the intracluster medium to well below 1 keV.
Nevertheless, photoelectric absorption by the cool gas remains an
important channel for information about its quantity and state, and
for this purpose it is important to be able separate the effects of
foreground absorption.

A number of authors have reported extreme ultraviolet (EUV) and soft
X-ray excesses in clusters of galaxies (e.g.  \citealt{lmb,klm};
\citealt*{bbk,blm}; but see \citealt{bb02}).  Such an excess could be
due to inverse Compton 
scattering of the microwave background by a non-thermal particle
population, but then the minimum particle pressure required in the
cluster Sersic 159-03 would exceed the thermal pressure of the hot gas
\citep{blm}.  In order to maintain hydrostatic equilibrium, that
would imply at least doubling the cluster mass.  Alternatively, the
soft excess in Sersic 159-03 may be due to gas that is cooler than the
hot intracluster medium, but then the total mass of cool gas would be
comparable to that of the hot gas \citep{blm}.  This presents the
problem that the huge mass of cool gas cannot be stably supported by
hydrostatic pressure.  Measurements of EUV and soft X-ray emission
from extragalactic sources are sensitive to correction for foreground
photoelectric absorption.  \citet{ab99} have argued on this basis
that the evidence for soft emission from a number of clusters is weak.
\citet*{kab} have argued further that arcminute scale holes in the
foreground absorption might account for at least part of the soft
excess.  So, the uniformity of foreground \hone\ emission is also of
interest for clusters showing EUV and soft X-ray excesses.

In section \ref{data} we describe the observations and data reduction.
Our main result is that there is no evidence for arcmin scale
structure in the foreground \hone\ emitting gas.  In section
\ref{discuss} we discuss the limits we can place on variation of
foreground \hone\ for each case.  For the purpose of obtaining
scales, we assume a flat $\Lambda$CDM cosmology, with $H_0 =
70\rm\, km\, s^{-1}\, Mpc^{-1}$ and $\Omega_{\rm m} = 0.3$ throughout
the paper.

\section[]{Observations and data reduction}\label{data}

We selected the cooling flow clusters A3581, A2597 and Sersic
159-03 because they have shown photoelectric absorption well in excess
of that expected due to foreground gas (\citealt{af97};
\citealt*{jft}).  As already noted, Sersic 159-03 is also reported as
having the greatest known soft X-ray excess \citep{blm}.  Pertinent
details of the clusters are given in Table~\ref{tbl:obs}.

\begin{table*}
  \begin{minipage}{160mm}
  \caption{Cooling flow clusters: observing and image parameters.}
  \label{tbl:obs}
  \centering
  \begin{tabular}{@{}lccccccccc@{}}
  \hline
  Cluster & \multicolumn{2}{c}{Position} & Redshift & Integration & \multicolumn{2}{c}{Natural weight}
  & \multicolumn{2}{c}{Uniform weight} \\
  & RA & Dec & & time\footnote{Total effective integration time after data
editing.} & Image beam\footnote{The dimensions of a Gaussian fitted to
the synthesized beam are given; they are mildly sensitive to the
image pixel size.} & RMS noise\footnote{The square root of the
pixel-to-pixel variance is given, calculated for regions of the the
image volumes devoid of sources.} & Image beam & RMS noise \\
  & \multicolumn{2}{c}{(J2000)} & & (hr) & (arcsec$^2$) & (mJy beam$^{-1}$) &
(arcsec$^2$) & (mJy beam$^{-1}$) \\
  \hline
  Abell 3581 & 14~07 & $-27$~01 & 0.023 & $8.6$ & $139 \times 51$ & $2.2$ & $85 \times 42$ & $3.5$ \\
  Sersic 159-03 & 23~14 & $-42$~43 & 0.058 & $10.3$ & $101 \times 77$ & $2.9$ & $60 \times 50$ & $5.0$ \\
  Abell 2597 & 23~25 & $-12$~06 & 0.085 &  $11.7$ & $396 \times 59$ & $2.9$ & $199 \times 43$ & $4.4$ \\
      \hline
    \end{tabular}
  \end{minipage}
\end{table*}

To image the Galactic foreground \hone\ which might be screening the
cooling flows at the centres of the clusters, we have used the
Australia Telescope Compact Array (ATCA) in two moderately compact
configurations: 375 and 750A.  The ATCA is an unfilled-aperture,
rotation synthesis interferometer, and as such measures only a finite
set of spatial (angular) components of the source distribution.  In
particular, the minimum baseline within the array sets the size of the
largest well-imaged structure, which for the configurations used is
$\sim10$~arcmin.  Features (ie.\ emission, absorption) in the source
distribution larger than this are largely invisible to the telescope.
To a good approximation then, our observations will produce images of
the ``fine structure'' in the foreground \hone, that is, the
modulation in the \hone\ above some pedestal value which varies slowly
across the sky and can only be measured by a filled-aperture (ie.\
single-dish) radio telescope.

Observations were acquired using the 375 configuration on 1996 December
4 and 7, and the 750A configuration on 1998 May 9, 10 and 12.
Observations of A3581 were made only with the 750A configuration.  The
half-power beam width of an ATCA antenna at wavelength 21~cm is
33~arcmin, and this sets the size of the field imaged by a single
pointing of the array; beyond $\sim 16$~arcmin from the field centre
the sensitivity falls off rapidly.  The image resolution within the
33~arcmin field is determined by the maximum antenna--antenna baseline
within the array, which for all fields was 735~m, corresponding to a
theoretical resolution of $\sim 60$~arcsec in the R.A. direction.
Because the ATCA is an East-West array and uses the Earth's rotation
to synthesize a planar interferometer, the resolution in the Dec
direction is elongated by a factor $\left| {\rm sin}^{-1}\delta
\right|$ for observations at Dec $\delta$; this effect is particularly
severe for A2597 at $\delta = -12^\circ06^\prime$.

To image Galactic gas, the ATCA receiver chain and correlator were
programmed to record XX and YY correlations between all telescope
pairings over a bandwidth of 8~MHz, centred on 1420~MHz.  The band was
divided into 1024 raw channels, yielding redshifted spectral channels of
width $1.65$~km~s$^{-1}$ extending from $-760$ to $+930$~km~s$^{-1}$
in the frame of the observations.\footnote{Note that the velocity {\em
resolution}\/ is $1.21$ times the channel width.}  The primary
calibrator 1934$-$638 was observed for absolute flux calibration
\citep{reynolds94} at the start of each observing period, and (weaker)
secondary calibrators were observed regularly during each observation
to enable the determination of time-dependent bandpass calibration
solutions.

The data were edited, calibrated and imaged in the radio
interferometry data reduction package {\sc miriad}.  Some editing of
the visibility data for the program sources was necessary where
on-line flagging had not been aggressive enough, and where transient
interference and/or system faults were evident.  Any continuum
emission present in the visibility data was removed by fitting and
subtracting a first-order polynomial from every visibility spectrum;
the fit was made to the outer channels in the band to avoid
contamination by Galactic {\sc Hi} line emission.  Radio continuum
images at 1420~MHz were made using the continuum fit data, and yielded
three sources with peak flux exceeding $10$~mJy in the A2597 and
Sersic 159-03 fields, and four in the A3581 field.  The clusters
themselves were detected at the following peak fluxes: A3581:
$0.40\pm0.02$~Jy; Sersic 159-03: $0.21\pm0.02$~Jy; and A2597:
$1.74\pm0.05$~Jy.

Images were produced by gridding and Fourier-inverting the
continuum-subtracted visibility data, averaging two raw channels at a
time and eliminating the (noisier) band edges, to produce images
having 200 planes (channels) of width $3.31$~km~s$^{-1}$, occupying
the radial velocity domain $-331 < V_{\rm LSRK} < +331$~km~s$^{-1}$
measured in the local standard of rest (kinematic) frame.  Image pixel
sizes were selected to sub-sample the synthesized beam by a factor of
$\sim 3$ in each direction.  We used two standard schemes to weight
the visiblity data during the inversion: uniform weighting which
assigns equal weight to all visibilities and produces images of high
angular resolution at the cost of increased noise; and natural
weighting, which favours visibility data from the shorter baselines
and produces images of lower angular resolution but significantly
lower noise.  The total effective integration time, image beam
dimensions and image noise levels for the observed fields are given in
Table~\ref{tbl:obs}.  The image noise levels are close to theoretical.
The spatial correlation scale in the images is well described by a
two-dimensional Gaussian having the dimensions listed in
Table~\ref{tbl:obs} (\ie\ the synthesized beam, typically three pixels
across).  Along the spectral axis, the images are $\ga 90$ per cent
uncorrelated.

\begin{figure}
\vspace{2.75in}
\includegraphics{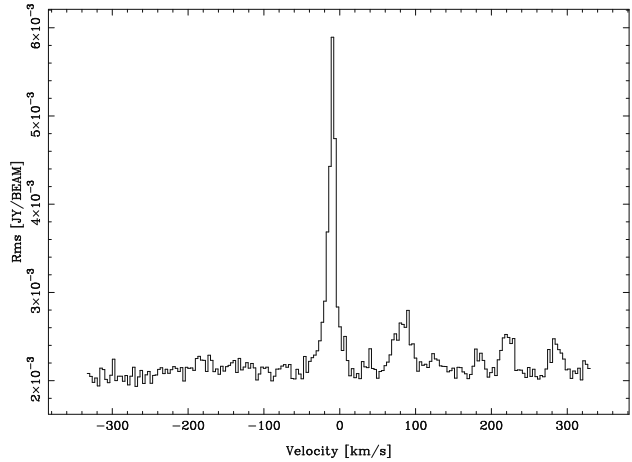}
\caption{RMS pixel variation per channel for A3581.  Features at -10
and 90~km~s$^{-1}$ are sidelobe artefacts of narrow linewidth sources well 
away from the primary beam of the telescope.  The features at 220 and
290~km~s$^{-1}$ are locations of genuinely increased noise probably
due to narrowband interference. \label{fig:rms1}} 
\end{figure}

\begin{figure}
\vspace{2.75in}
\includegraphics{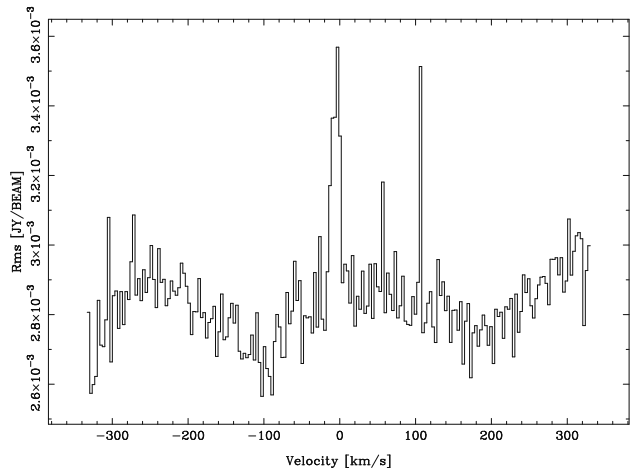}
\caption{RMS pixel variation per channel for Sersic 159-03.  The
slightly elevated noise level near 0~km~s$^{-1}$ is due to a faint
grating pattern across the channel maps, probably arising from
internal interference. \label{fig:rms2}}
\end{figure}

\begin{figure}
\vspace{2.75in}
\includegraphics{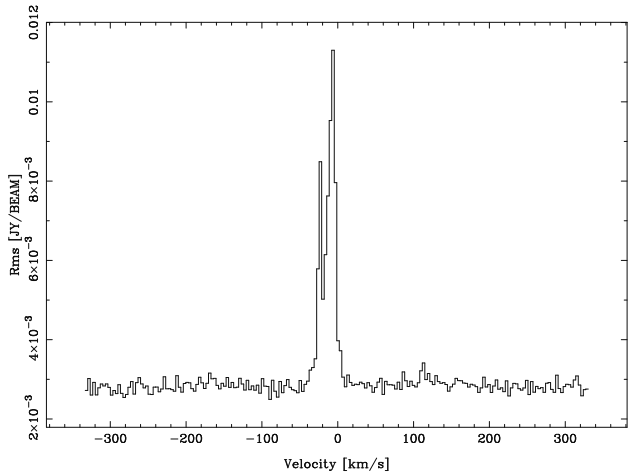}
\caption{RMS pixel variation per channel for A2597.\label{fig:rms3}} 
\end{figure}

Table~\ref{tbl:obs} lists image noise values for regions of each image
which are free of obvious \hone\ emission sources.  To assess any
variation in image noise across the field, we produced maps of the RMS
pixel value along each sky pixel in the non-primary-beam-corrected
image cubes.  Towards each cluster, the RMS varied by less than
10~per~cent from the value listed in Table~\ref{tbl:obs} over the sky.
To assess any varation in image noise through the frequency space of
the observations, we produced a spectrum of the RMS pixel variation
per channel; the RMS noise spectra are reproduced in
Figures~\ref{fig:rms1}--\ref{fig:rms3} for the
naturally-weighted images.  The spectra again show very little
variation throughout the frequency domain, except for a few features
which we now discuss.

\paragraph*{A3581.}  The features at $-10$ and $+90$~km~s$^{-1}$ in the
RMS spectrum towards A3581 (Figure~\ref{fig:rms1}) are due to
sidelobes of strong, narrow linewidth and point-like sources well
outside the primary beam of the telescope.  Careful inspection of the
calibrator spectra rules out the chance that these are artefacts
introduced by Galactic absorption towards the secondary calibrator
source, and consequently we believe them to be real ``hot spots'' in
the Galactic \hone\ column.  These features are similar to the
signature this project aims to find, but are well away ($>40$~arcmin)
from the line-of-sight to the cluster, and are simply bright enough to
be detected in the sidelobes of the telescope.  The features at $220$
and $290$~km~s$^{-1}$ are channels having genuinely increased noise
due to narrowband interference, some of which has been flagged.  Apart
from these features, the noise in this field is steady at
$2.2$~mJy~beam$^{-1}$.

\paragraph*{Sersic 159-03.} This field exhibits slightly elevated noise
near 0~km~s$^{-1}$, where a grating pattern is evident in the
channel maps.  This is either due to interference (probably internal
to the system) which is too weak to identify and flag in the raw
visibility data, or to a strong, narrow source passing through a
distant sidelobe of the telescope.  It is certainly not characteristic
of local emission (or absorption) along the line of sight to the
cluster.  It only marginally affects our ability to detect a
foreground excess of the magnitude required (see below).  The noise in
this field is generally less than $2.9$~mJy~beam$^{-1}$.

\begin{figure*}
\vspace{4.3in}
\includegraphics{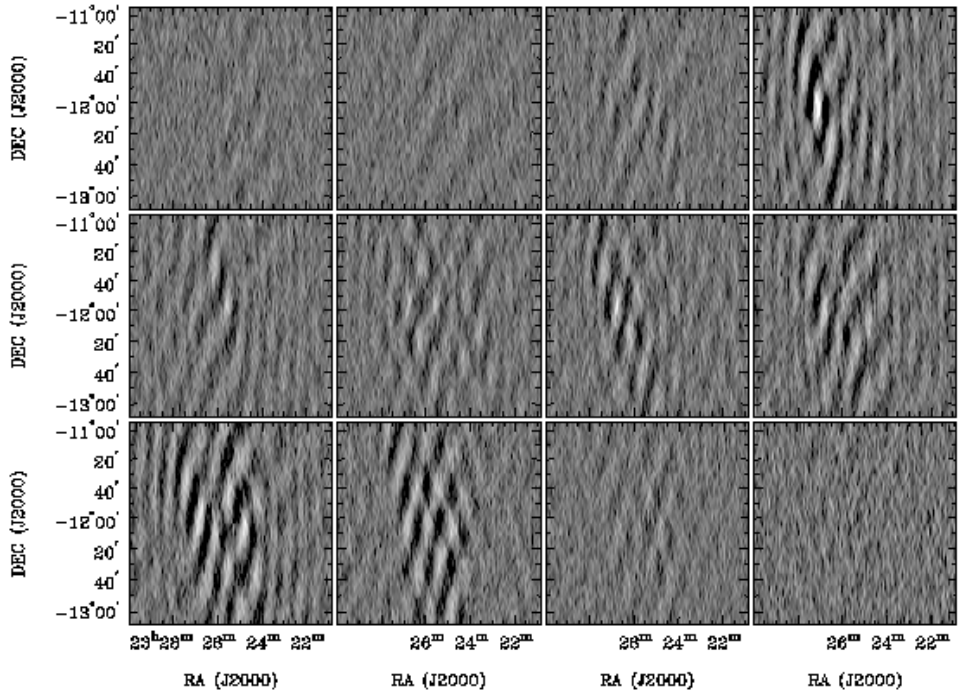}
\caption{Raw channel maps (ie.\ un{\sc clean}ed) near 0~km~s$^{-1}$
for A2597.  The greyscale extends from $-10$ to $+30$~mJy~beam$^{-1}$.
Sequence is left-to-right, top-to-bottom, with the map at
top-left at LSR velocity $-33.9$~km~s$^{-1}$, and bottom-right
$+1.7$~km~s$^{-1}$.  \label{fig:map}} 
\end{figure*}

\paragraph*{A2597.}  The \hone\ emission in the A2597 field is complex
and difficult to interpret.  Figure~\ref{fig:rms3} shows substantial
``pollution'' extending from $-30$ to $0$~km~s$^{-1}$, of an otherwise
flat noise profile.  In Figure~\ref{fig:map} we present individual
channel maps covering this velocity range, selected from the
naturally-weighted image.  In these un{\sc clean}ed maps, sources are
convolved with the synthesized beam and so even a simple source
distribution on the sky could produce the bulk of the observed
pattern; elliptical sidelobes are particularly evident in the
bottom-left and top-right panels of Figure~\ref{fig:map}.  However
there are also fainter grating patterns, most obvious in the second
and third panels, but possibly extant in all the maps.  Grating
patterns are generally the result of a fixed interference source
present for some or all of the observations and despite our efforts,
it has not been possible to find and eliminate the offending data in
this case.  

\begin{figure*}
\vspace{4.3in}
\includegraphics{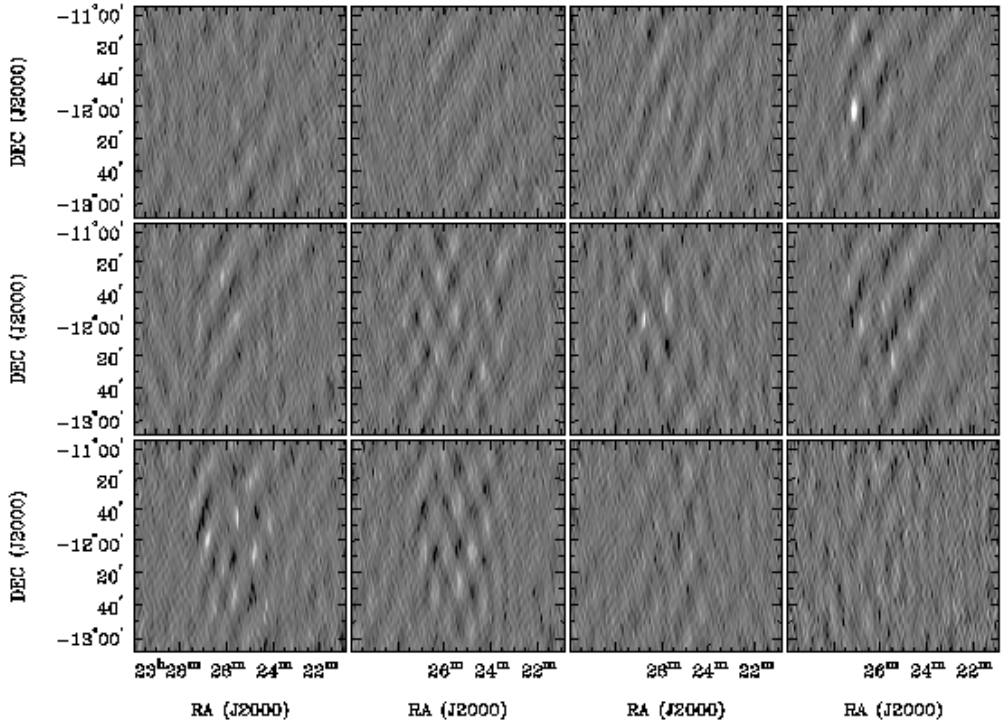}
\caption{{\sc Clean}ed channel maps for A2597.  Details otherwise
identical to Figure~\ref{fig:map}. \label{fig:restor}} 
\end{figure*}

In Figure~\ref{fig:restor}, we show the result of iteratively
deconvolving the channel maps of Figure~\ref{fig:map} and restoring
the images with a Gaussian beam fitted to the synthesized beam, using
the {\sc miriad} tasks {\sc clean} and {\sc restor}.  The resultant
maps, aside from the genuine hot-spot in the top-right panel, and
possibly those in the bottom-left panel (none of which are aligned
with the cluster cooling flow), appear to contain a semi-regular array
of sources.  It seems far more likely that these are artefacts where
multiple grating patterns reinforce, and reanalysis confirms this: by
producing images with a different selection of antenna pairings and/or
weighting schemes, the synthesized beam sidelobe structure can in
principle be modified such that an array of sources on the sky would
produce a less degenerate raw map, but such a combination of antennae
was not found.  Forthwith, for our analysis we simply consider the
A2597 data to comprise a base noise level of $2.9$ ($4.4$)
~mJy~beam$^{-1}$ and an elevated noise level of $10$ ($15$)
~mJy~beam$^{-1}$ in the regime $-30$ to $0$~km~s$^{-1}$ for naturally
(uniformly) weighted images.  

\begin{table*}
  \begin{minipage}{160mm}
    \caption{Limits on arcmin scale variations in foreground neutral
      hydrogen column density}
    \label{nhlimits}
    \centering
    \begin{tabular}{@{}lccccccccc@{}}
      \hline
      Cluster& \multicolumn{2}{c}{${\rm N}_{\rm HI}$} & Channels\footnote{number of $3.3\rm\, km
      \,s^{-1}$ channels with significant foreground emission in HIPASS
      data.} & \multicolumn{3}{c}{Natural weight} &
      \multicolumn{3}{c}{Uniform weight}\\
      & S92\footnote{\cite{stark92}, $\sim 2$~degree beam} &
HB97\footnote{\cite{hartmann97}, $\sim 35$~arcmin beam}  & & \multicolumn{3}{c}{$\Delta {\rm N}_{\rm HI}\ (10^{20}\, \psqcm)$} 
      & \multicolumn{3}{c}{$\Delta {\rm N}_{\rm HI}\ (10^{20}\, \psqcm)$} \\ 
      & \multicolumn{2}{c}{$(10^{20}\, \psqcm)$} & &
      summed\footnote{$5\sigma$ noise per channel summed over the
      channels with foreground emission.} & averaged\footnote{summed
      $5\sigma$ noise divided by the square root of the number of
      contributing channels.} & {\bf measured}\footnote{image-plane RMS in
moment map generated over HIPASS-determined channels} & summed
&averaged &{\bf measured} \\ 
      \hline
      Abell 3581 &4.5 & 4.1 &36 &2.0 &0.34 & {\bf 0.62} &6.4 &1.1 & 1.2\\
      Sersic 159-03 &1.8 & n/a &64 &4.4 &0.55 & {\bf 0.78} &17 &2.1 & 2.5 \\
      Abell 2597 &2.5 & 2.5 &48 &1.1 &0.16 & {\bf 0.43} &5.1 &0.74 & 0.66 \\
      
      \hline
    \end{tabular}
  \end{minipage}
\end{table*}

\section[]{Results and discussion}\label{discuss}

We calculate upper limits to arcmin scale variations in the
foreground neutral hydrogen column density towards the cooling flows
in three different ways.  The results are listed in
Table~\ref{nhlimits}.  Columns headed ``summed'' present the $5\sigma$
image noise per channel (as given in Table~\ref{tbl:obs}) multiplied
by the number of channels in which foreground emission is evident in
{\sc Hi} Parkes All Sky Survey spectra\footnote{We use
HIPASS spectra to estimate the likely spectral extent of the
foreground column, as it is a single-dish survey which sees emission
at (nearly) all angular scales on the sky.} \citep[HIPASS,][]{b01}, and
converted to a column density in the standard way, ie.\
\begin{equation}
{\rm N}_{\rm HI} = 1.36 \cdot \left( 1.823 \times 10^{18} \right) \cdot
        {{\lambda^2} \over {\theta_a \theta_b}} \cdot {\rm S}_{\rm int},
\end{equation}
where the factor $1.36$ incorporates the sensitivity of the ATCA,
$\lambda$ is the observing wavelength (cm), $\theta_{a,b}$ are the
major and minor widths of the synthesized observing beam (arcsec,
listed in Table~\ref{tbl:obs}), ${\rm S}_{\rm int}$ is the summed flux
(Jy beam$^{-1}$ m s$^{-1}$), and the result is in atoms cm$^{-2}$.
Summed limits on beam-to-beam fluctuations in column density are given
for both the natural and uniform weight maps in Table~\ref{nhlimits},
illustrating the practical trade off between spatial resolution and
noise level.

The summed calculation gives a very conservative upper limit to the
beam-to-beam fluctuations in ${\rm N}_{\rm H}$.  These limits apply
even if the fluctuation in every channel is in the same sense.
However, as noted above, there is little or no correlation from
channel to channel in the image slices, so that the contributions of
the channels to the noise are largely independent.  In this case, the
noise in the summed image will be the RMS value for the combined
channels.  If the channels also contribute equally to the noise, the
RMS noise equals the summed noise divided by the square root of the
number of channels.  Upper limits on foreground column density
calculated in this way are given in Table \ref{nhlimits} under the
heading ``averaged.''

Finally, in the columns headed ``measured'' of Table~\ref{nhlimits},
we list a third set of upper limits to the foreground neutral column
density.  For each field, we collapse the channel maps by summing
those channels for which HIPASS spectra show emission to be present
(ie.\ we generate a zeroth-order moment map) and then calculate the
RMS fluctuations in the image plane of the moment map.  Importantly,
this technique takes into account the channel-to-channel image noise
variation, shown in Figures~\ref{fig:rms1}--\ref{fig:rms3}.  In all
cases, these ``measured'' column density limits lie between the
``average'' and ``summed'' values, and we believe them to be the most
appropriate limits to use in further analysis.

For comparison with our results, in columns~2 and 3 of
Table~\ref{nhlimits} we list the foreground {\sc Hi} column
interpolated from \cite{stark92} and \cite{hartmann97} (where
available); these data are averaged over beams of area $\sim 4.5$ and
$\sim 0.4$ square degrees respectively.  Our adopted $5\sigma$ limits
(bold) on the fluctuations in the foreground column are considerably
less than the wide-area averaged column itself for the natural weight
images (beam areas $\sim 2$--7 square arcmin), but are on the order of
the \cite{stark92} column for our uniform weight images (beam areas
$\sim 1$--2 square arcmin).  Therefore, our key result is that the
foreground \hone\ emission is structureless over the range of scales
probed by our natural weighted observations for all three clusters.
This outcome is consistent with the extrapolation of the \hone\ power
spectra measured at larger angular scales by
\cite{green93}, and with the agreement in \hone\ column
measured by the \cite{stark92} and \cite{hartmann97} surveys whose
beams differ by an order of magnitude in area. Comments on the
individual clusters follow.

{\bf Abell 3581} is a moderately poor cluster \citep{pg92} and,
perhaps, the least well studied of our objects.  Its central galaxy,
IC 4374, is a cD galaxy hosting the radio source PKS 1404$-$267.
Based on {\it ASCA} and {\it ROSAT} data, \citet{jft} found a mass
deposition rate of $80 \, \msunpyr$, and an absorbing column density
of $2\times10^{21}\, \psqcm$, well in excess of the foreground value
(Table \ref{nhlimits}).  Using a redshift of 0.0222
\citep[corresponding to the recession velocity of $6567\rm\, km\,
s^{-1}$;][]{jft}, the scale is $0.45\rm\, kpc\, arcsec^{-1}$, so that
our naturally weighted beam size is $62 \times 23$ kpc at the cluster.
This is well inside the cooling flow region, which is about 100 kpc in
radius \citep{jft}.  An excess column density of $2\times10^{21}\,
\psqcm$ would violate our $5\sigma$ limit if it covered a disc
exceeding 5 kpc in radius projected onto the cluster.  It is very
unlikely that foreground Galactic gas could account for the excess
absorption in this object.  Our $5\sigma$ limit on fluctuations in
foreground
\hone\ column density from beam-to-beam is about 14~per~cent.  This
constrains the properties of any holes in the foreground gas, as
discussed below.

{\bf Sersic 159-03}, or Abell S1101, has been observed with {\bf
  XMM-Newton} \citep{kft} and the results show that very little gas
cools below $\sim 1$ keV.  This undermines earlier claims of excess
photoelectric absorption in this cluster \citep{af97}.  At the
redshift of the cD galaxy, $z = 0.0564$ \citep{mdw}, the cluster scale
is $1.1\rm\, kpc\, arcsec^{-1}$, so that our naturally weighted beam
size is $110 \times 84$ kpc.  The $5\sigma$ limit on fluctuations per
beam in the foreground \hone\ column density is 43~per~cent for this
cluster.  Note that, as discussed in the Introduction, Sersic 159-03 
is the cluster with the greatest reported soft X-ray excess
\citep{blm}.  

{\bf Abell 2597} has been observed extensively at a wide range of
wavelengths.  It shares many of the noteworthy properties of cooling
flow clusters, including a central radio source, PKS 2322$-$122
\citep{sbr}, optical emission lines \citep{vd97}, recent star
formation \citep{kos} and spatially extended cool gas \citep{dmv,ewj}.
\citet{afj} found significant excess photoelectric absorption, but
{\it Chandra} observations show cavities \citep[``ghost'' cavities
that do not surround the currently detectable radio lobes;][]{mwn},
and that the bulk of the gas does not cool below $\sim 1$ keV.  However,
FUSE data do show evidence of some gas cooling to lower temperatures
\citep{ocd}.  At a redshift of 0.0821 \citep{vd97}, the cluster scale
is $1.6\rm\, kpc\, arcsec^{-1}$, making our naturally weighted beam
size $610 \times 91$ kpc.  Our $5\sigma$ limit on the foreground
\hone\ fluctuation per beam is 17~per~cent for this source.

It is clear that foreground \hone\ emitting gas has little to do with
excess photoelectric absorption in any of the three cooling flow
clusters considered here.  The lack of structure on arcminute scales
in any of these clusters tells us that such structure is uncommon at
best, at least for sources that are well clear of the Galactic plane.

In the light of the EUV and soft X-ray excesses, for the remainder of
this section we focus on placing constraints on holes in the
foreground gas \citep{kab}.  Our data only place useful constraints on
the properties of holes that are comparable in size to the radio beam.
For the purpose of discussion, we consider a simple model, with the
sky divided into small areas, $a$, typical of the size of a hole, and
smaller than the area $A$ of the radio beam.  These areas are covered
at random with column density $N_0$ or $N_1 < N_0$, with the smaller
column density to represent holes.  If the fraction of the sky
covered by holes is $f$, then the average column density will be
$\langle {\rm N}_{\rm HI} \rangle = N_0 - f (N_0 - N_1)$.  The number of hole
areas falling within the radio beam has a binomial distribution, and
the RMS fluctuation in the measured column density is $\sigma_N = 
(N_0 - N_1) \sqrt{f(1-f)a   /A}$.  The fractional fluctuation in
column density from beam-to-beam is then 
\[
\delta = {\sigma_N\over \langle {\rm N}_{\rm HI} \rangle } = {\Delta
  \sqrt{f(1-f)a/A} \over 1 -   f\Delta},
\]
where $\Delta = (N_0 - N_1)/N_0$.

If holes in the foreground \hone\ distribution are to account for EUV
and soft X-ray excesses, then the column density in the holes must
be significantly lower than average, \ie\ we need $\Delta \simeq 1$, so
we will take this to be exact.  The limits on fluctuations in \hone\
column density from Table \ref{nhlimits}  may then be regarded as 
constraining the size of the of any holes, \ie 
\[ 
\sqrt{a\over A} < \delta \sqrt{1 - f \over f},
\]
where $\delta$ is now the observed limit on the fractional fluctuation
in ${\rm N}_{\rm HI}$.  For example, if the covering fraction of the holes
is 1/2, then the constraint is $\sqrt{a/A} < \delta$, giving
upper limits to the angular diameter of any holes of $14$, $46$ and
$32$ arcsec for Abell 3581, Sersic 159-03 and Abell 2597,
respectively, in order to comply with the limits in Table
\ref{nhlimits}.  It is unfortunate that our weakest constraint applies
to Sersic 159-03, the cluster with the greatest reported soft X-ray
excess \citep{blm}.

It is interesting to consider what is required of the distribution of
the \hone\ gas in order to produce a significant covering fraction of
holes.  Here, the phrase ``low density gas'' is used loosely to refer
to gas which, if spread along a line of sight, would appear as a
significant hole in the foreground \hone\ screen.  

The average fraction of a random line of sight that lies in low
density gas is equal to the filling factor of 
that gas.  In order to appear as a hole, the bulk of the line of sight
must pass through low density gas.  Thus, if the covering fraction of
holes is substantial, the filling factor of the low density gas needs
to be close to unity.  The most plausible distribution of \hone\ gas
that could appear smooth down to arcmin scales, yet still produce a
substantial covering fraction of holes is one where the gas is highly
clumped on small scales.  The clumps must be significantly smaller
than one arcmin in order for the 21 cm emission to appear smooth on
that scale.  At distances comparable to the gas scale height, $\sim
100$ pc, a 1 arcmin clump would be $\sim 0.03$ pc in size.  On the
other hand, in order to produce appreciable small-scale fluctuations
in ${\rm N}_{\rm HI}$ a typical line of sight can only encounter a small
number of clumps.  This means that the column density of one clump
must be comparable to the mean value of ${\rm N}_{\rm HI}$.  For small,
roughly spherical clumps this requires very high gas densities, $n \ga
10^3\rm\ cm^{-3}$, almost certainly exceeding the limits on the gas
pressure.  This constraint is relaxed somewhat if the clumps are
really filaments, or folds in sheets \citep{dl90}.

Some studies have found variations in foreground \hone\ column density
on scales as small as 10 milliarcsec (\eg\ \citealt*{ddg};
\citealt{fg01}).  However, these variations are seen only in the
coldest \hone\ component, and are by no means normal: \cite{fg01}
report that only two of seven background continuum sources illuminate
opacity gradients in the foreground column.  \cite{deshpande00} offers
a compelling argument that the overall impact on the \hone\ column
along different lines of sight from {\em physically allowable}\/
clouds of cold
\hone\ is likely to be very small, and this is again consistent with
the studies of the \hone\ power spectrum on larger angular scales
\citep[\eg][]{green93}.

\section[]{Conclusions}

Radio maps of the 21 cm emission from foreground Galactic neutral
hydrogen over 3 cooling flow clusters show that the gas is uniform on
scales from $\sim 1$ to 10 arcmin.  Our $5\sigma$ limits on
beam-to-beam fluctuations in 21 cm emission, expressed in terms of
limits on the fluctuations in \hone\ column density, are
$6.2\times10^{19}$, $7.8\times10^{19}$ and $4.3\times10^{19}\ \psqcm$
for the clusters Abell 3581, Sersic 159-03 and Abell 2597,
respectively, on scales of the order of 1 arcmin (see Table
\ref{nhlimits}).  Variations in the column density of neutral hydrogen
in the Galaxy do not account for the claimed excess photoelectric
absorption in these clusters.

In the light of recent detections of significant excess soft X-ray
emission from some clusters, we have also considered whether ``holes''
in the distribution of foreground neutral hydrogen could cover a
significant fraction of the sky, and can confidently rule out a high
covering fraction of holes at scales $\sim 1$ arcmin.

\section*{Acknowledgments}

The Australia Telescope is funded by the Commonwealth of Australia for
operation as a National Facility managed by the CSIRO.  We thank Shane
Isley, Mirjam Jonkman and Vincent McIntyre for their help in obtaining
the ATCA data.  We gratefully acknowledge Lister Staveley-Smith for
his advice on reducing the data.  We thank Rachel Webster and Martin
Meyer for comments on the manuscript.  We thank the referee for a
thorough and constructive report.

\label{lastpage}

\end{document}